# The solar neutrino project HELLAZ : status report on the hardware and the simulation.

J. Dolbeau, on behalf of the HELLAZ Collaboration
*Physique Corpusculaire et Cosmologie - Collège de France*
*11 place Marcelin Berthelot  F-75005 Paris*
E-mail: dolbeau@cdf.in2p3.fr

## Abstract

We report the status of the solar neutrino project called HELLAZ. In a high pressure and cooled TPC filled with helium, the detection of each ionisation electron of the track produced by the recoil electron from elastic scattering induced by solar neutrinos, will yield the low energy, high resolution solar neutrino spectrum. Two topics are presented: tests of Micromegas gas chambers giving a good discrimination between adjacent ionisation electrons and the preliminary test of the direction of the primary electron track from the cloud analysis. These two tests seem rather encouraging.

HELLAZ is a rather ambitious project aimed at measuring the solar neutrino spectrum with a threshold as low as possible (close to 100 keV) and a high resolution (5%). The reaction used is the elastic neutrino-electron scattering, known very accurately over the full energy range. The main aim is to separate the $^7$Be line from the pp spectrum, and also, according to the shape of the pp spectrum to be able to decide which kind of eventual oscillation is taking place.

The final detector shall be a time projection chamber (TPC) of 2000 m$^3$, (10 m diameter and 20 m long, divided in two: the maximum drift being 10 m) filled with He at 10 bar and cooled to 150K. The choice of helium is dictated by the facts that the electron drift velocity is low, the electron diffusion is low and it is possible, through boil-off to produce the gas totally immune from radioactive contaminants. The choice of the low temperature comes from the requirement that the radon should be frozen so that no radioactive source moves in the gas and makes any subtraction impossible. The target is then 2•10$^{30}$ electrons, ensuring a yield, if all neutrinos from the standard solar model are detected, of about 4 $^7$Be and 7 pp neutrino induced events per day with the low foreseen threshold (100 keV electron energy, that is 200 keV neutrino energy). The recoil electron makes a track (5 to 50 cm) of ionisation electrons which drift at 1.4 mm/μs under the influence of an electric field (200 V per cm) towards the 2D end cap gas detector. The track is reconstructed in 3D and knowing its energy and angle relative to the sun, one is able to calculate the neutrino energy. Previous reports can be found in [1]. We show here the status on two parts of the R&D.

**Tests with Micromegas chambers**

The neutrino energy resolution requires that the electron energy is measured with a resolution better than 5%, which is not too difficult, and that the recoil direction of the electron has to be determined to within 2°. This last requirement is difficult. As a matter of fact, the common use of ADCs to measure the total charge falling on a wire or a strip of the end cap gaseous detector will yield somewhat inaccurate results for such low energy tracks. We aim here to gather the maximum possible information, and for that we plan to measure each individual ionisation electrons. A track drifting at 1.4 mm/μs contains about 200 electrons per cm. The diffusion coefficient $S_D$ has been measured [2]. Its value in micrometers is:

$$S_D = 200\sqrt{L_D}$$

where $L_D$ is the drift length expressed in centimeters. As a matter of fact, this coefficient is measured in the transverse direction and it is assumed up to now that the longitudinal coefficient equals the transverse. This has still to be checked. In average, electrons are separated by some 5 mm. The first incidence is that the X and Y strips of the detector (we cannot use pads for economic reasons in a 20 m$^2$ surface) should not be narrower than 2 mm. The second incidence is that the electrons arrive on these strips with a waiting lines law, that is in $e^{-t/\tau}$ as a time function, the function being linear in the strip width. We immediately see that the shortest the pulses of the gas detector, the more efficient the system will be to count all electrons one by one.

After having tried different designs, we have chosen the Micromegas chamber [3]. In this design, the field is constant in the gap between anode and cathode, contrary to the 1/r field around the wire in a usual wire chamber. So, first, the avalanche is spread over a wider distance, which makes higher gains more easy.

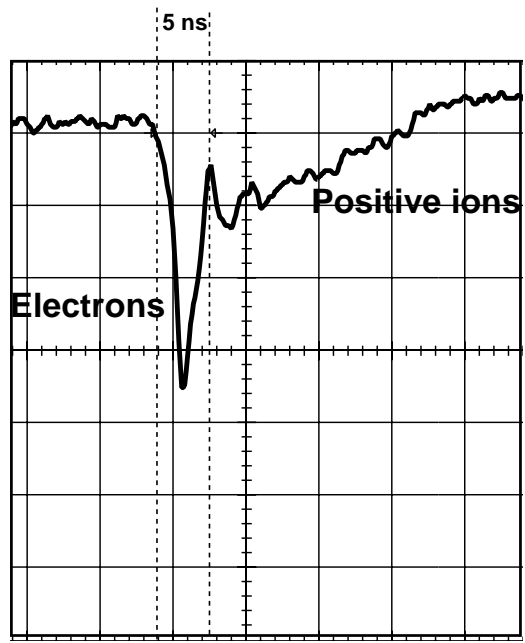

Figure 1 : A typical Micromegas pulse seen by a fast current preamp, showing the fast and slow components.

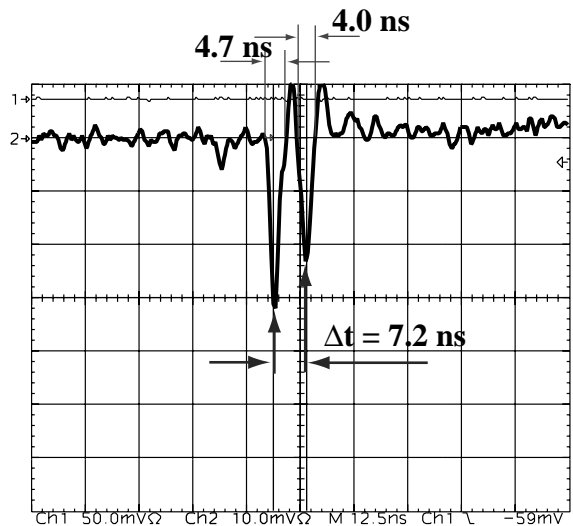

Figure 2: Two pulses separated by 7.2 ns. Pressure is 1 bar He, 15% $CH_4$, 100μm gap

Second, the electrons produced in the avalanche travel a longer distance to the anode, and, in a slow gas like He, the signal due to the electrons movement is one to two ns long, and so can be seen with a fast current preamp (Ortec, [4]). Illuminating a Ni-Cr photocathode with the light of a 250 nm laser, propagated through 2 fibers of different length, one is able to see 2 pulses separated in time [5]. With such an apparatus, at one atmosphere and room temperature, we show fig. 1 a pulse where the 2 components (electrons and positive ions) can clearly be seen. Fig. 2 shows 2 pulses (filtered with a 50 pf capacitor at the preamp output to remove the slow components) separated by 7 ns. Finally, we show fig. 3 the 0 and 1 photoelectron spectrum recorded by looking at the amplitude (and not the charge) of the fast pulse. The preamp is here a cheap home made «common base» designed to replace the Ortec one. The single electron efficiency here is about 98%.

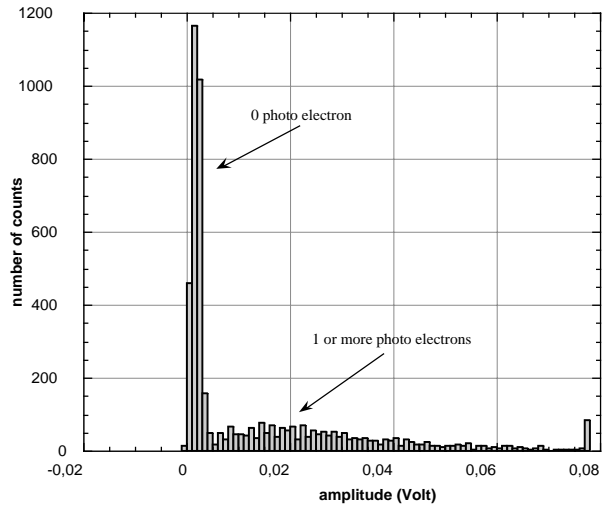

Figure 3: Single electron spectrum (corresponding to fig.2 conditions), recorded by amplitude analysis.

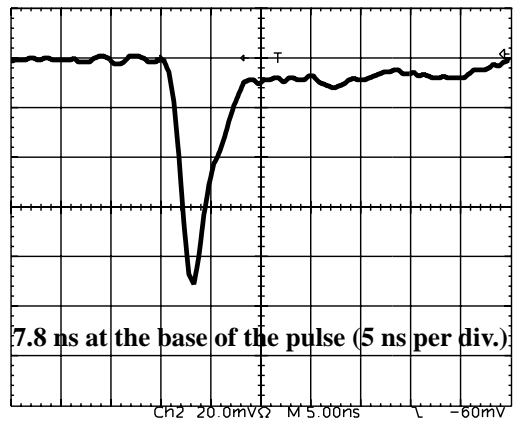

Figure 4: Typical Micromegas pulse: 10 bar He, 2% $CH_4$, 50 µm gap

The pressure in our vessel is now 10 bar. The mean free path for the propagation of the avalanche is much shorter so that the gap is now 50 µm. We have also decreased the quantity of $CH_4$ quencher to a few %. Fig. 4. shows a typical pulse: its width at the base is still well under 10 ns. We have not yet reached a gain such that we can see single electrons and this is due to a few problems we are in the process of solving: first, the mesh (cathode) suffers a pressure towards the anode about 10 times higher, so we have to control the mesh tension. Second, the cleanness of the vessel becomes more critical and we have to mount our detectors in a clean room.

Until now, Micromegas chambers at our disposal are only monodimensional. We have a simple scheme to produce X-Y chambers with an anode printed circuit deposited on double face kapton. This is based on two interleaved «combs». The size of the comb teeth has to be analogous to the avalanche spread. We are now measuring that quantity.

**Reconstruction of the electron track**

The precision on the angle of the reconstructed track is the second crucial quest of the experiment feasibility. Extreme caution must be taken, and verifications have to be done before publishing a meaningful result . We have not reached that point yet, and what follows is more a description of the actual state of the simulations (which algorithms have been followed) than a final result: all along what follows, the points towards which a big effort has to be made in order to arrive at a credible conclusion will be stretched.

The simulations are made with GEANT 3.21. The very first point to check is the quality of the tracks simulations: the energies are very low (one hundred to a few hundred keV) and it is maybe one of the very first times where the precise behaviour of electrons in a high density gas is simulated to be checked experimentally.

The tracks analysis, prior to any drift, shows clearly that the best fit is the power 3/2 approximation in a plane projection. If the electron is emitted along the z axis, the orthogonal deviations x and y can be expressed by:

$$x = a\, z^{3/2}$$
$$y = b\, z^{3/2}$$

This is true only when the length of the trajectory tends towards zero. This law is easy to understand intuitively: the charged particle will deviate from a straight trajectory of a quantity x (resp. y) proportional to the length z and to the square root of the number of atomic hits. Practically, this stays a good approximation for a length varying from 5 to 20 mm when the electron energy varies from 100 to 400 keV. The precision ($\sigma_\theta$) if the interaction vertex is known, is around 1° (@ $E_e \geq$ 300 keV) and 4° (@ 100 keV). The corresponding values for a linear fit are respectively 2° and 5°. However, with the supposition that transverse and longitudinal diffusions are equal, we can see that after 7 m of drift, the transverse distance between electrons will be around 1 cm, and so roughly the same as the length of the initial track on which the power 3/2 fit is performed. The memory of the initial track is thus seriously degraded: this leads to the use of a longer initial track, including probable «accidents» and the lowest the initial energy is, the more difficult the problem will be.

An algorithm has been built which allows, by successive approximations to get a good estimation of the interaction point also called the vertex (sigma is 1-2 mm for a drift of 4 m ).

In a first step, one tries to reconstruct the track from the cloud, with an ideal end-cap gas detector. The useful cloud zone around the vertex can be determined in 98% of the cases (tracks of 300 keV with 1 m drift) to 75% (100 keV and 4 m drift). Most of the miss should be recuperated with a more refined algorithm.

In a second step, under study, one simulates the real gas detector at the end cap to estimate the losses and the first step is re-done with the effectively measured electrons.

To conclude on the state of advancement of the track angle reconstruction program, we show the distributions of the angles between the projections on a plane of the true direction and the reconstructed direction in the case of 500 electrons of 300 keV and a drift of 100 cm (fig. 5). A gaussian fit yields an uncertainty less than 5° (for 7 tracks, the vertex is clearly not found and only some 10 tracks have problems - angle > 20°). These uncertainties are presented fig. 6 as a function of the drift length between 0 and 800 cm for 3 energies: 100, 200 and 300 keV. These results have been produced by linear fits instead of the power 3/2 fits which are slightly worst, contrary to what was observed with zero drift and a known vertex. Because the power 3/2 fit is extremely sensitive to the vertex position reconstruction, this tells us that we do not yet reconstruct the vertex with a sufficient accuracy, for the diffusion introduced by the drift does not change the shape of the initial electron track due to elastic scattering, it just washes out the irregularities.

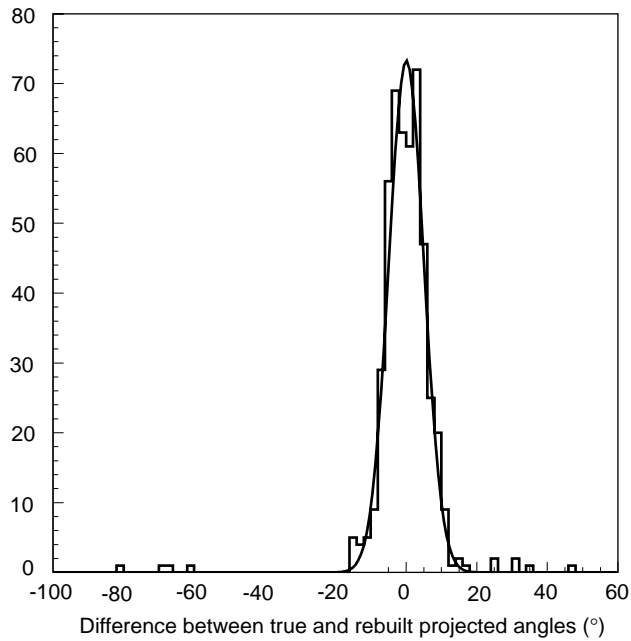

Figure 5 : Distribution of the direction of the reconstructed electron (300 kev, 100 cm drift). The sigma of the gaussian (here 4.97°) is what is reported vertically in fig. 6.

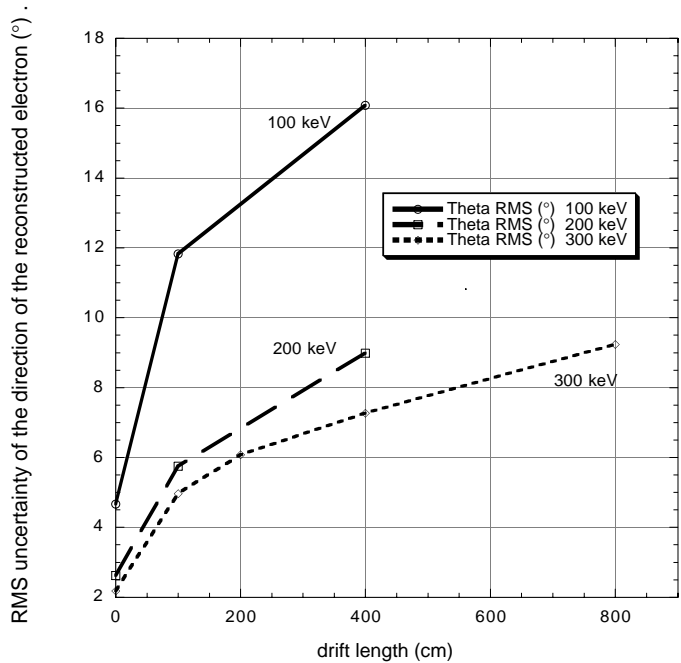

Figure 6 : It is easier to reconstruct the direction of a «high» energy electron.

In the near future, we plan to create and measure real low energy electron tracks in a small TPC (25 cm drift and 100 cm$^2$ detector surface) and inject them in the reconstruction program.